\documentclass [preprint] {revtex4}

\usepackage[dvips]{graphicx}
\usepackage{amssymb}

\def\lp {\left( }
\def\rp {\right) }
\def\lb {\left[ }
\def\rb {\right] }
\def\lc {\left\{ }
\def\rc {\right\} }
\def\ra {\rangle }
\def\la {\langle }

\def\rar {\rightarrow}
\def\lrar {\leftrightarrow}

\def\beq{\begin{equation}}
\def\eeq{\end{equation}}
\def\bea{\begin{eqnarray}}
\def\eea{\end{eqnarray}}
\def\nn {\nonumber}

\def\vs {\vspace}

\def\Ob {\bar{\Omega}}

\def\Kb {\bar{K}}

\def\rtw {\sqrt{2}}

\def\sp {\!+\!}
\def\sm {\!-\!}

\def\cK {{\cal{K}}}

\def\a{\alpha}
\def\b{\beta}
\def\d{\delta}

\def\m{\mu}
\def\n{\nu}

\def\p {\pi}
\def\r{\rho}

\def\T {\Theta}

\def\bpi {\mbox{\boldmath $\pi$}}

\begin{document}

\title{electro-weak $\bpi\bpi$ form factor and $\bpi\bpi$ scattering: 
\\ 
towards a phenomenological tool}

\author{M. R. Robilotta}
\affiliation{ Instituto de F\'{\i}sica, Universidade de S\~{a}o Paulo,  
S\~{a}o Paulo, SP, Brazil}
 
\date{\today }

\begin{abstract}

The weak two-pion form factor $F_V^{\p\p}$ is described 
as the product of a weak kernel $\cK_W$ by a strong function $\T_{\p\p}^P$, 
determined directly from $\p\p$ scattering data.
As the latter accounts at once for all effects associated with resonances,
intermediate $K\Kb$ loops, and other possible inelasticities present in $\p\p$ scattering,
the need of modeling is restricted to $\cK_W$ only.
The  procedure proposed allows one to asses the weak kernel directly,
which has a dominant cut beginning at the $K\Kb$ threshold.
Even the simplest vector-meson-dominance choice for $\cK_W$ already yields
a good qualitative description of $F_V^{\p\p}$. 
The energy sector below $0.8$ GeV is quite well reproduced
when a precise theoretical  chiral perturbation $\p\p$ amplitude
is used as input, together with the single free parameter $F_V G_V/F^2=1.20$.
The inclusion of kaon loops, along well established lines and using few parameters, 
produces a good description of the form factor in the entire energy range allowed by $\tau$ decays.
This indicates that the replacement of modeling by direct empirical 
scattering information can also be useful in the construction  of  theoretical 
tools to be used in analyses
of hadronic heavy meson decay data.

\end{abstract}

\pacs{...}

\maketitle

\section{motivation}

In the last decade, a considerable amount of precise data has been
produced on the form factor $F_V^{\p\p}$,
measured in the reactions $e^+ e^- \rar \p\p$
and $\tau \rar \p\p \n$\cite{Cleo,Aleph,Belle}.
This motivated a corresponding theoretical effort, aimed at explaining
the features found, which involve a rich hadronic final 
state structure\cite{th}.

Final state interactions of the same kind  also intervene in some $D$ and $B$ decays. 
They may be more visible in $\p\p$ semi-leptonic modes,
but also contribute to $\p\p\p$ and $K\p\p$ final states,
as well as to $KK\p$ and $KKK$ reactions, through
$\p\p\sm K\Kb$ couplings.
At present, the large amount of data produced by the 
LHCb facility at CERN is allowing the precise identification
of a wide variety of multi-meson final states,
which is instrumental for studying CP violation\cite{IT}.
On general grounds, theoretical descriptions of hadronic decays 
involve two distinct sets of interactions.
One of them concerns the primary weak vertex, in which a heavy quark,
either $c$ or $b$, emits a $W$ and becomes a $SU(3)$ quark.
As this process occurs within the heavy meson, 
it amounts to the effective transition of a $D$ or a $B$ into 
a first set of $SU(3)$ mesosns.
This is followed by purely hadronic final state interactions, 
in which the mesons produced in the weak decay rescatter
before being detected.
As both weak and final state interactions include several 
competing processes, the treatment of heavy-meson decays 
into hadrons is necessarily involved.

As the extraction of data from Dalitz plots is complicated,
experimental analyses tend to rely on relatively simple guess functions,
which do not distinguish primary weak vertices from final state
interactions.
In fact, most groups employ just the so called isobar model, 
based on sums of different Breit-Wigner functions,
somehow inspired by free meson-meson structures.
However, at present, data acquired at LHCb have very high statistics,
accurate partial wave analyses are available, and existing guess
functions prove to be no longer effective\cite{AI}.
This unsatisfactory situation imotivates the present work.
In order to move forward,  
one needs to understand how weak and strong interactions 
become entangled in decays containing two final mesons. 

The process $\tau \rar \p \p \n$ is possibly the simplest laboratory
for this kind of study, since energies available are relatively high,
the leptonic sector is well understood, 
and high-quality data exist for the  form factor $F_V^{\p\p}$.
As one is dealing with a two-pion decay,  
an important part of hadronic final state interactions is encoded into the elastic $\p\p$ amplitude.
In principle, consistent theoretical models must deal simultaneously with these two observables and,
as consequence,  several parameters are required for good  fits of both data sets. 
In the present work, one shows that it is possible to eliminate the elastic $\p\p$ amplitude from the problem,
by describing  the form factor $F_V^{\p\p}$
as the product of a weak kernel $\cK_W$ 
by a strong function, determined directly from  scattering data.
This restricts both modeling an fitting efforts to just $\cK_W$, 
which can be reliably approximated by a real function up to 1 GeV.

In the next Sect., one isolates the weak kernel $\cK_W$, relying mostly on topological arguments,
in Sect. III, tree-level fits are tested and, in Sect. IV, a simple model is presented, which yields a 
satisfactory description of $|F_V^{\p\p}|$ data.
Sect.V contains conclusions and, for completeness,  technical matters
are summarized into three short Appendices.

\section{vertex $W \rar \p\p$}

\begin{figure}[h] 
\includegraphics[width=1\columnwidth,angle=0]{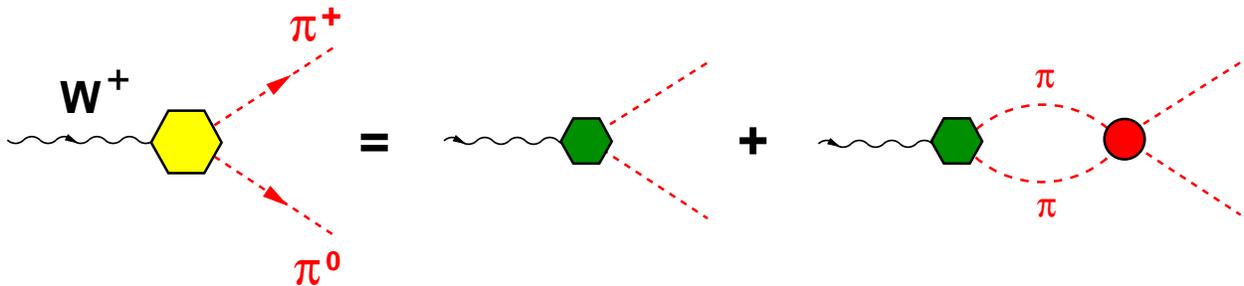}
\caption{Structure of the $W \rar \p \p$ matrix element:
the small green hexagon and the red blob represent, respectively, 
the weak kernel $\cK_W$ and the on-shell elastic $\p\p$ amplitude $T_{\p\p}^P$.}
\label{FV}
\end{figure}

The dynamic structure of $F_V^{\p\p}$ is determined by a primary 
weak kernel and subsequent strong final state interactions.
Since this kind of separation is not born in nature, but part of a theoretical strategy,
there are many possibilities for defining these two complementary sets.
The two-pion system is prominent in this problem and it is convenient to 
isolate the on-shell $\p\p$ elastic amplitde $T_{\p\p}^P$.
Therefore, one defines the weak kernel  $\cK_W$ as the set 
encompassing all diagrams attached to the weak vertex which cannot be cut along
two on-shell pion lines only.
This gives rise to a structure for the vertex $W\rar \p\p$ as in Fig.\ref{FV},
where the kernel $\cK_W$,  represented by the small green hexagon,
has cuts only above the $\p\p$ threshold.
This decomposition allows the part of $F_V^{\p\p}$ associated with
the $\p\p$ amplitude to be treated in a model independent way.

%

The two-pion weak vertex is parametrized in terms of the form 
factor $F_V^{\p\p}$ as
\bea
&& \la \p^+(p_1) \, \p^0(p_2) |V^\m|0\ra =
\rtw \; F_V^{\p\p}(s)\; (p_1 \sm p_2)^\m \;,
\label{1}
\eea
with $s=(p_2 \sp p_1)^2$ and, in the isospin limit, 
this matrix element is exactly conserved.
As both $\cK_W$ and $T_{\p\p}^P$ have internal substructures,
it is convenient to organize their contributions to $F_V^{\p\p}$
in terms of kernels and amplitudes.
The former are represented by $\cK$ and correspond to diagrams
which cannot be split into two parts by cutting 
two on-shell meson lines only,
whereas the latter, denoted by $T$, include kernels iterated
by means of intermediate two-meson propagators $\Ob$.
In the case of $P$-waves, both kernels and amplitudes 
are proportional to $(t \sm u)$ and 
one writes $\cK=(t \sm u)\,\cK^P$ and $T=(t \sm u)\,T^P$.

The treatment of  coupled channel amplitudes is 
traditional and has already been employed in ref.\cite{Hyams}.
The basic kernels $\cK_{ij}^P$ are defined as:
$\cK_{11}^P \rar [\p\p \lrar \p\p]$,
$\cK_{12}^P \rar [\p\p \lrar KK]$,
$\cK_{22}^P \rar [KK \lrar KK]$.
Together with the two-meson propagators $\Ob_{ii}^P$, 
given in App.\ref{propagator}, they are used to construct 
amplitudes, as discussed in App.\ref{kernels}.
The $\p\p$ amplitude can be cast 
in the usual form
\bea
&&T_{\p\p}^P = 
\frac{\cK_{11}^P 
+ \Ob_{KK}^P ||\cK||}
{1 + \Ob_{\p\p}^P \, \cK_{11}^P + \Ob_{KK}^P \, \cK_{22}^P
+ \Ob_{\p\p}^P\,\Ob_{KK}^P \,||\cK||} \;.
\nn\\[2mm]
&& ||\cK|| = \lb \cK_{11}^P\,\cK_{22}^P \sm (\cK_{12}^P)^2\,\rb \;,
\label{2}
\eea
which is equivalent to the  more compact expression
\bea
&& T_{\p\p}^P = \frac{\cK_{\p\p}^P}{1 + \Ob_{\p\p}^P \,\cK_{\p\p}^P} \;,
\nn\\[2mm]
&& \cK_{\p\p}^P = \cK_{11}^P 
- \cK_{12}^P \lb 1 \sm \Ob_{KK}^P\, T_{22}^P \rb
\Ob_{KK}^P \, \cK_{12}^P \;,
\label{3}
\eea
where $\cK_{\p\p}^P$ becomes complex above the $KK$ threshold.

The structure of $F_V^{\p\p}$,  as given in Fig.\ref{FV}, reads
\bea
F_V^{\p\p}(s) =
\cK_W(s) \, \lb 1 - \Ob_{\p\p}^P(s) \; T_{\p\p}^{P}(s) \rb \;,
\label{4}
\eea
where $\cK_W$ is the weak kernel, to be discussed in 
the sequence.
This result clearly indicates that the form factor cannot
be directly proportional to the scattering amplitude,
as assumed in many analyses of heavy-meson decay data.
Using eq.(\ref{3}), it can be rewritten as
\bea
&& F_V^{\p\p} = \cK_W \, \T_{\p\p}^P \;,
\label{5}\\
&& \T_{\p\p}^P = \lb \frac{1}{1 + \Ob_{\p\p}^P\,\cK_{\p\p}^P} \rb \;.
\label{6}
\eea
The function $\T_{\p\p}^P$, already noticed in Ref.\cite{BR}, can be related directly to 
the free $\p\p$ amplitude, as discussed in App.\ref{empf}.
It is determined just by the empirical phase shift $\d$
and inelasticity parameter $\eta$, as
\bea
&& \T_{\p\p}^P = \frac{1}{1 - i \, \tan \theta }
\nn\\[2mm]
&& i\, \tan\theta = \frac{[\eta^2 -1] + i\,[2 \eta \sin 2\d ]}
{1+\eta^2 + 2 \eta \cos 2\d} \;.
\label{7}
\eea
This is important because, being derived from experiment,
this function accounts automatically for all effects associated with 
pion loops, resonances,
intermediate $K\Kb$ loops, and other possible inelasticities present in $T_{\p\p}^P$.
This sector of the problem becomes then model independent.

\begin{figure}[h] 
\hspace*{-30mm}
\includegraphics[width=.7\columnwidth,angle=0]{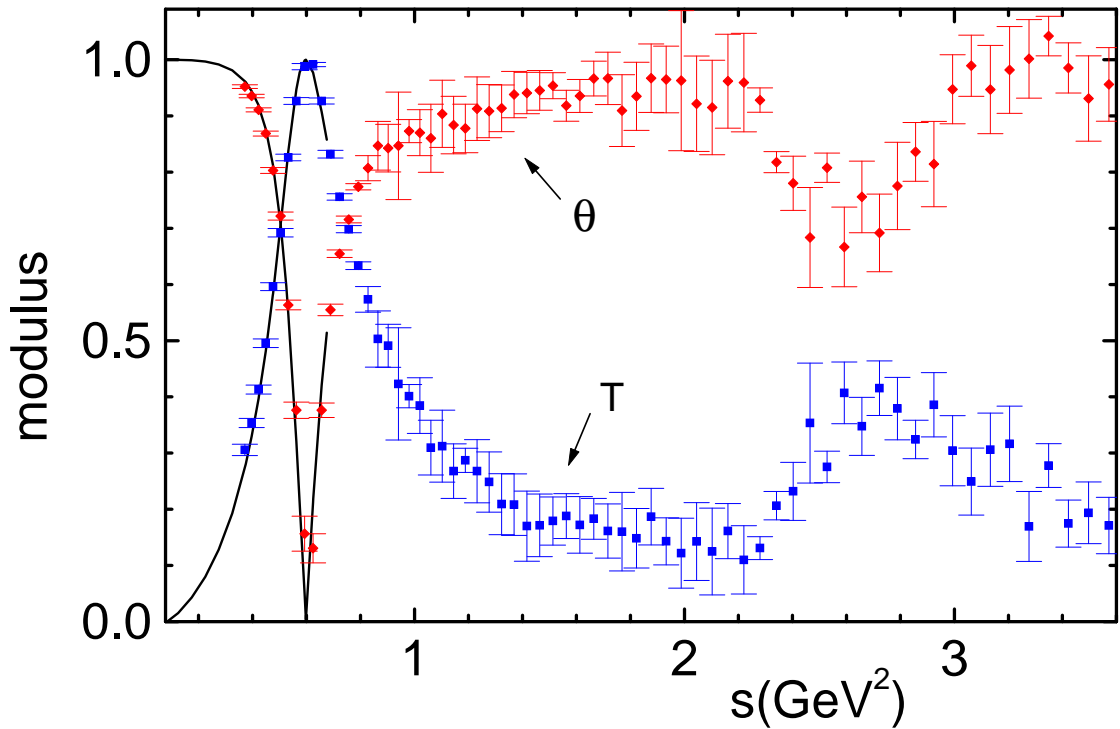}
\caption{Moduli of the elastic $P$-wave $\p\p$ amplitude (T, blue) and
of the function $\T_{\p\p}^P$, eq.(\ref{7}), ($\theta$, red)
based on data from Ref.\cite{Hyams}. 
The continuous curves (black) were obtained from the chiral analysis
of Ref.\cite{CGL} and are valid up to $0.65\,$GeV$^2$.}
\label{TheTpp}
\end{figure}
%

In Fig.\ref{TheTpp}, one shows the modulus of the elastic 
$P$-wave $\p\p$ amplitude, obtained from both CERN-Munich data\cite{Hyams}
and the low-energy analysis by Colangelo, Gasser and Leutwyller
(CGL)\cite{CGL}, which holds up to $0.65\,$GeV$^2$,
together with the corresponding predictions 
for $|\T_{\p\p}^P|$, given by Eq.(\ref{7}).
One notes that the latter departs from $1$ at threshold 
and displays resonances as dips, instead of bumps.
This feature may prove to be instrumental 
to data analyses of other decay processes involving
two-pion final states.

\begin{figure}[h] 
\hspace*{-30mm}
\includegraphics[width=.7\columnwidth,angle=0]{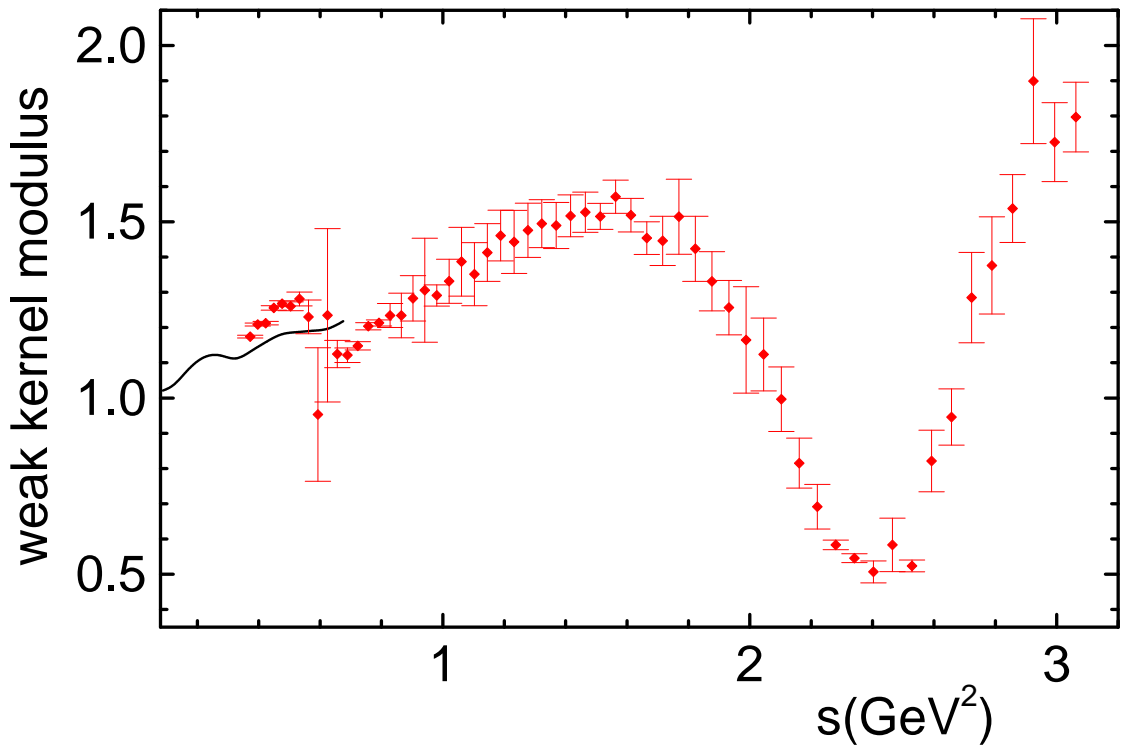}
\caption{Modulus of the weak kernel, eq.(\ref{5})multiplied
by $[1-s/m_\rho^2]$, using either data from Ref.\cite{Hyams}
(red squares) and the chiral analysis of Ref.\cite{CGL} 
(continuous black line), which is valid up $0.65\,$GeV$^2$.}
\label{KW}
\end{figure}

As both $F_V^{\p\p}$ and $\T_{\p\p}^P$ can be extracted from experiment,
Eq.(\ref{5})  allows one to isolate direct empirical information about $\cK_W$. 
One finds, in Fig.\ref{TheTpp}, that $\T_{\p\p}^P$ vanishes at the $\rho$-pole,
whereas tree-level models for 
$\cK_W$ involve a bare $\rho$ propagator.
This suggests that it is convenient to consider the regular combinations
$[\cK_W/F_\rho]$ and $[F_\rho \, \T_{\p\p}^P]\,$, where
$F_\rho = 1/[1-s/m_\rho^2]$ is the usual vector-meson-dominance (VMD) form factor\cite{GL, EGPR}.
The ratio $|\cK_W/F_\rho|$,
obtained by using Belle data\cite{Belle}
together with the $\T_{\p\p}^P$ from Fig.\ref{TheTpp},
is given in Fig.\ref{KW}.
It oscillates around the $\rho$-pole, 
owing to numerical mismatches between different
choices for  $m_\rho$ in the inputs used, 
has a roughly linear growth from threshold up to $s=1.8\,$Gev$^2$
and a pronounced dip, with a minimum around $2.4\,$GeV$^2$.

%

\section{tree-level results}
\label{weakmodel}

\begin{figure}[h] 
\hspace*{-30mm}
\includegraphics[width=.7\columnwidth,angle=0]{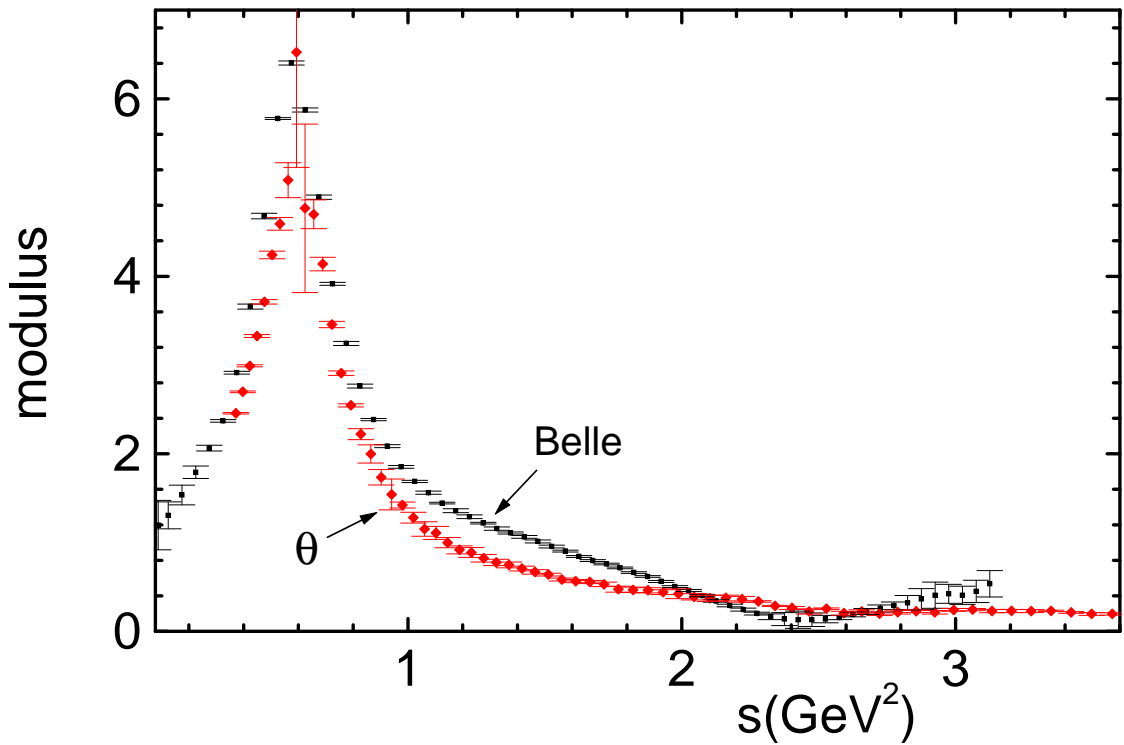}
\caption{Moduli of the Belle\cite{Belle} form factor (Belle, black) 
and the factor $|\Theta_{\p\p}^P/(1-s/m_\rho^2)|\,$, eq.(\ref{7}),
($\theta$, red), based on data from Ref.\cite{Hyams}.}
\label{FVThe}
\end{figure}

The leading tree contributions to $\cK_W$ read
\bea
&& [\cK_W]_{tree} = \lb 1 
- \frac{F_V\,G_V}{F^2} \, \frac{s}{s \sm m_\r^2} \rb \;,
\label{8}
\eea
using the notation of Ref.\cite{EGPR}.
The choice $F_V\,G_V/F^2=1$ yields the usual VMD expression
and corresponds to the minimal prediction for $F_V^{\p\p}$, 
displayed in Fig.\ref{FVThe}, together with Belle data\cite{Belle}.
These curves are interesting because they show that
$F_\r \T_{\p\p}^P$  is largely dominant for $s<1$ GeV$^2$ and still quite significant
over the entire range of energies considered.
This indicates the extent of the overlap between $F_V^{\p\p}$ and 
$\p\p$ effects encoded into final state interactions.
Moreover, the relatively small deviations between both sets of points provide a model 
independent indication of missing weak kernel structures.

%

One considers the low-energy region in Fig.\ref{MOa}, 
and compare  Belle data\cite{Belle} with predictions based on the chiral $\p\p$ amplitude
by Colangelo, Gasser and Leutwyller( CGL)\cite{CGL}, 
using the value $F_V\,G_V/F^2=1.20\,$ in Eq.({\ref{8}),
which ensures the coincidence with the datum for $|F_V^{\p\p}|$ at the
closest point to the $\rho$-mass.
This choice is not far from the determination made in Ref.\cite{SCP}.
The agreement between Belle points and predictions from the CGL chiral analisis,
which relies on a single free parameter,  is quite good.

\begin{figure}[h] 
\hspace*{-30mm}
\includegraphics[width=.7\columnwidth,angle=0]{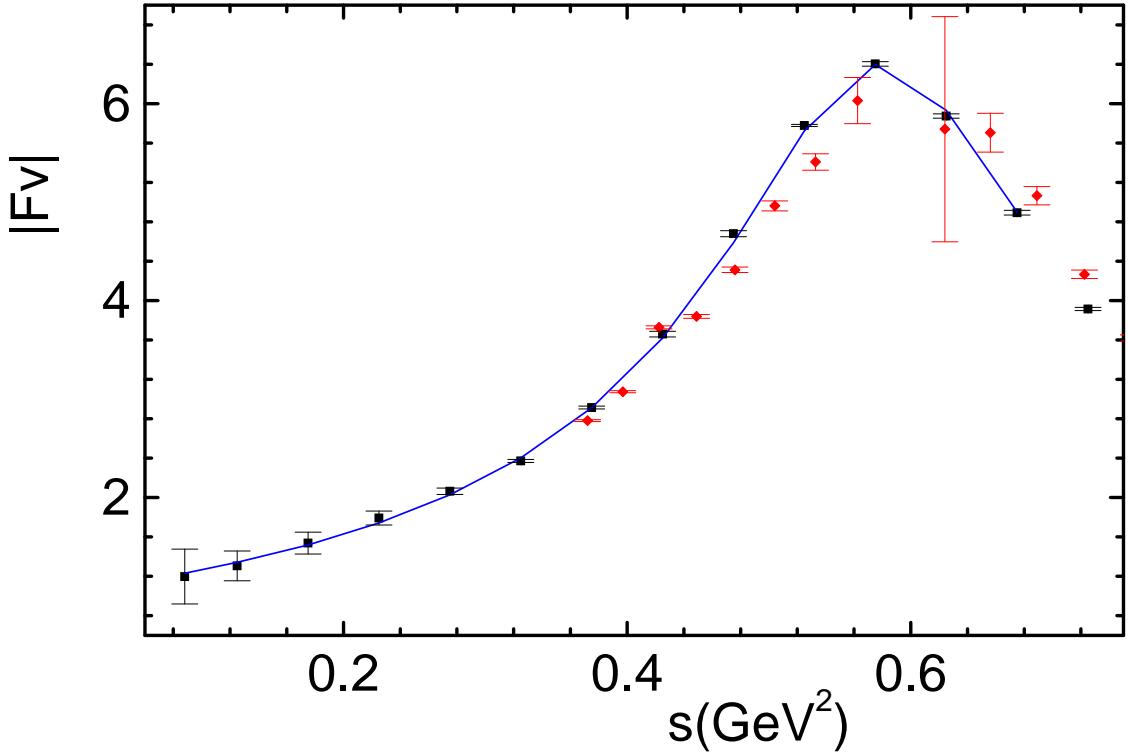}
\caption{Moduli of the Belle\cite{Belle} form factor (black)
and model results using CERN-Munich data \cite{Hyams} (red)
and the chiral analysis of Ref.\cite{CGL} (contnuous blue curve);
the curve was drawn by using CGL results
at energies where data exists and interpolating them by 
means of straight lines, just to guide the eye.}
\label{MOa}
\end{figure}
%

\section{a simple model}
\label{weakmodel}

\begin{figure}[h] 
\includegraphics[width=1\columnwidth,angle=0]{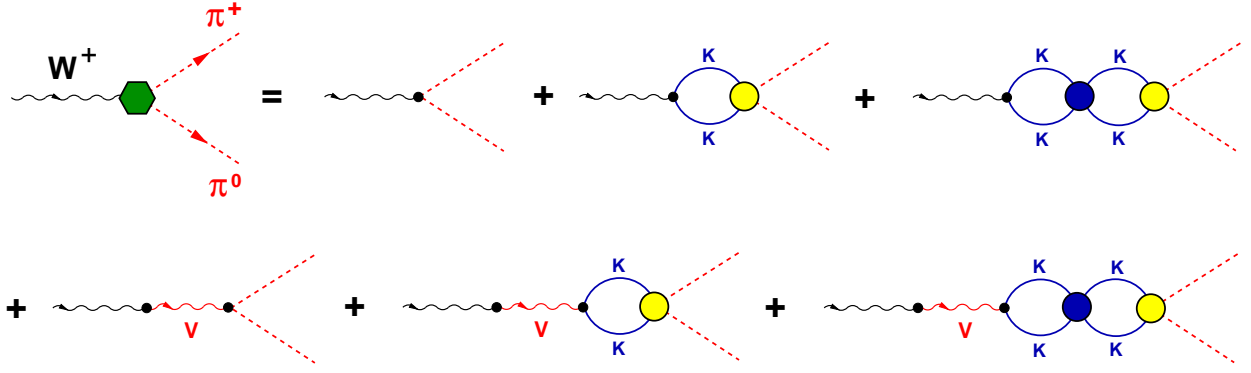}
\caption{Model for the weak kernel, involving kaon loops
and both $\rho$ and other vector mesons.}
\label{model}
\end{figure}
%

In this work, one is interested in the structure of final state $\p \p$
interactions which occur in a broad class of weak decays, and
does not  aim at a very precise description of $|F_V^{\p\p}|$,
which could compare with those produced by experts\cite{th}.
In the region $s>1$ GeV$^2$, one considers just a simple model for $\cK_W$,
suggested by  Fig.\ref{KW} and given in Fig.\ref{model}, which includes well known tree-level\cite{GL, EGPR} 
and coupled channel interactions. 
Contributions from loops involving virtual pions, 
assumed to be small, are excluded.
Using the conventions of Ref.\cite{EGPR} 
and two vector mesons, the tree term is given by 
\bea
&& [\cK_W]_{tree} = \lb 1 
- \frac{F_V\,G_V}{F^2} \, \frac{s}{s \sm m_\r^2} 
- \a \, \frac{s}{s \sm m_v^2} \rb \;,
\label{9}
\eea
where $\a\,$ and $m_v$ are free parameters. 
This allows the four kaon-loop terms of Fig.\ref{model} to be written as 
\bea
[\cK_W]_{loop} &\!=\!& -\,\frac{1}{\rtw}\; [\cK_W]_{tree}\;
\nn\\
&\!\times \!&
\lb 1 - \Ob_{KK}^P \, \frac{\cK_{22}^{P}}
{1 + \Ob_{KK}^P \,\cK_{22}^P} \rb \;
\Ob_{KK}^P \, \cK_{12}^P
\nn\\[2mm]
&\!=\!& -\,\frac{1}{\rtw}\;[\cK_W]_{tree}\;
\lb \frac{\Ob_{KK}^P \, \cK_{12}^P}
{1 + \Ob_{KK}^P \,\cK_{22}^P} \rb \;
\label{10}
\eea
The weak kernel is given by the sum of eqs.(\ref{9}) and (\ref{10})
and reads
\bea
[\cK_W]_{model} &\!=\!& [\cK_W]_{tree}\;
\lb 1 - \frac{1}{\rtw} \,
\frac{\Ob_{KK}^P \; \cK_{12}^{P1}}
{1 + \Ob_{KK}^P \,\cK_{22}^P} \rb \;.
\label{11}
\eea
The $SU(3)$ relation $\cK_{12}^P= \rtw\,\cK_{22}^P$ yields
\bea
&& [\cK_W]_{model} = [\cK_W]_{tree}\; \T_{22}^P \;, 
\nn\\[2mm]
&& \T_{22}^P =\lb \frac{1} {1 + \Ob_{KK}^P \,\cK_{22}^P}  \rb \;.
\label{12}
\eea
As there are no data for $\cK_{22}^P$,
one assumes a form inspired by 
chiral symmetry\cite{GL, EGPR} 
\bea
\cK_{22}^P= \frac{1}{2 F^2} \lb 1 - \b \,\frac{s}{s \sm m_v^2}\rb \;,
\label{13}
\eea
where $\b$ is another free parameter.
This, together with the two-kaon propagator, eq.(\ref{A5}),
determines the model, given by the product of eq.(\ref{13}) by  the empirical $\T_{\p\p}^P$,  which reads
\bea
&& [F_V^{\p\p}]_{model} = 
[\cK_W]_{model} \; \T_{\p\p}^P
\nn\\[2mm]
&& = \lb 1 
- \frac{F_V\,G_V}{F^2} \, \frac{s}{s \sm m_\r^2} 
- \a \, \frac{s}{s \sm m_v^2} \rb \; \T_{22}^P \; \T_{\p\p}^P \;.
\label{14}
\eea

One compares the model  with Belle data\cite{Belle},
by determining $\T_{\p\p}^P$ from 
CERN-Munich $\p\p$ results\cite{Hyams}  and 
keeping the  choice $F_V\,G_V/F^2=1.20\,$, made earlier.
The other thee parameters, namely $\a=0.01081\,$,
$\b= 0.79374\,$, and $m_v=1.5491\,$GeV
were obtained by fitting central values of C-M data 
to interpolated central Belle data.
This procedure for fitting a data set to another one
is definitely not precise and is aimed just at producing
a visual feeling for the model properties.
Complete result is given in Fig.\ref{MO} and,
given the crudeness of the model,
the overall agreement can be taken as satisfactory.
Results at the high-energy end are shown in Fig.\ref{MOb},
where one notes both a small mismatch around $1.9\,$GeV$^2$
and a zero for the model, just above $2.6\,$GeV$^2$.

\begin{figure}[h] 
\hspace*{-30mm}
\includegraphics[width=.7\columnwidth,angle=0]{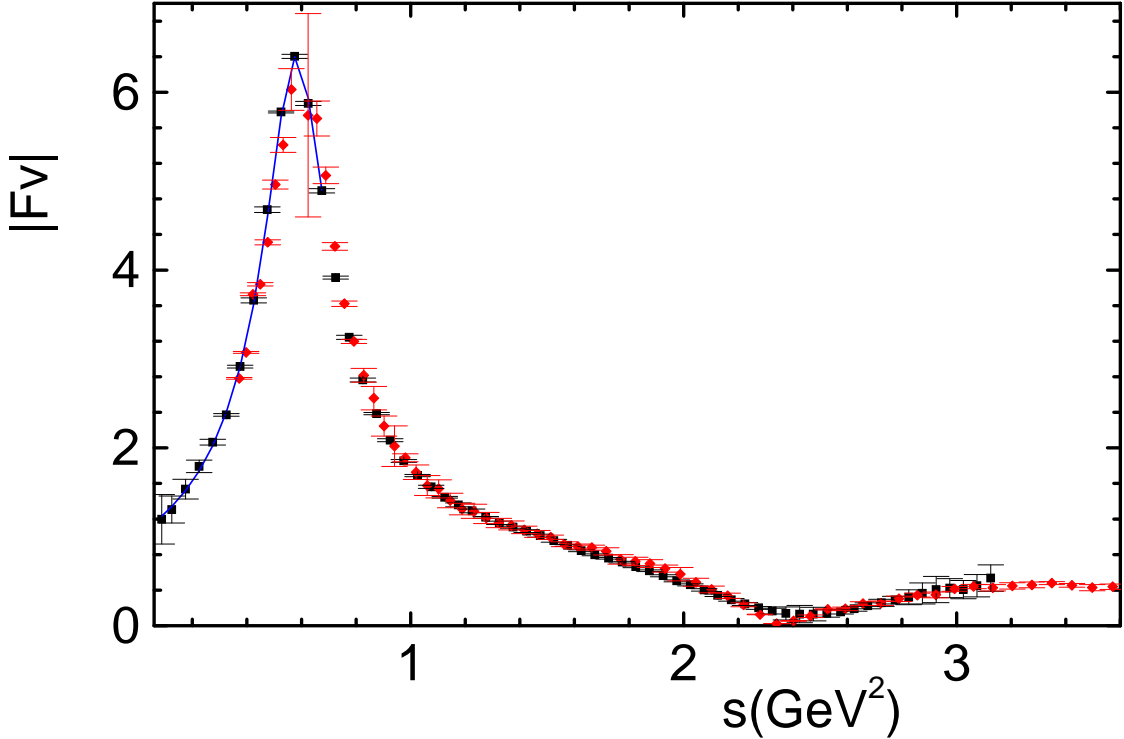}
\caption{Moduli of the Belle\cite{Belle} form factor (black)
and model results using CERN-Munich data \cite{Hyams} (red)
and the chiral analysis of Ref.\cite{CGL} (contnuous blue curve).}
\label{MO}
\end{figure}

\begin{figure}[h] 
\hspace*{-30mm}
\includegraphics[width=.7\columnwidth,angle=0]{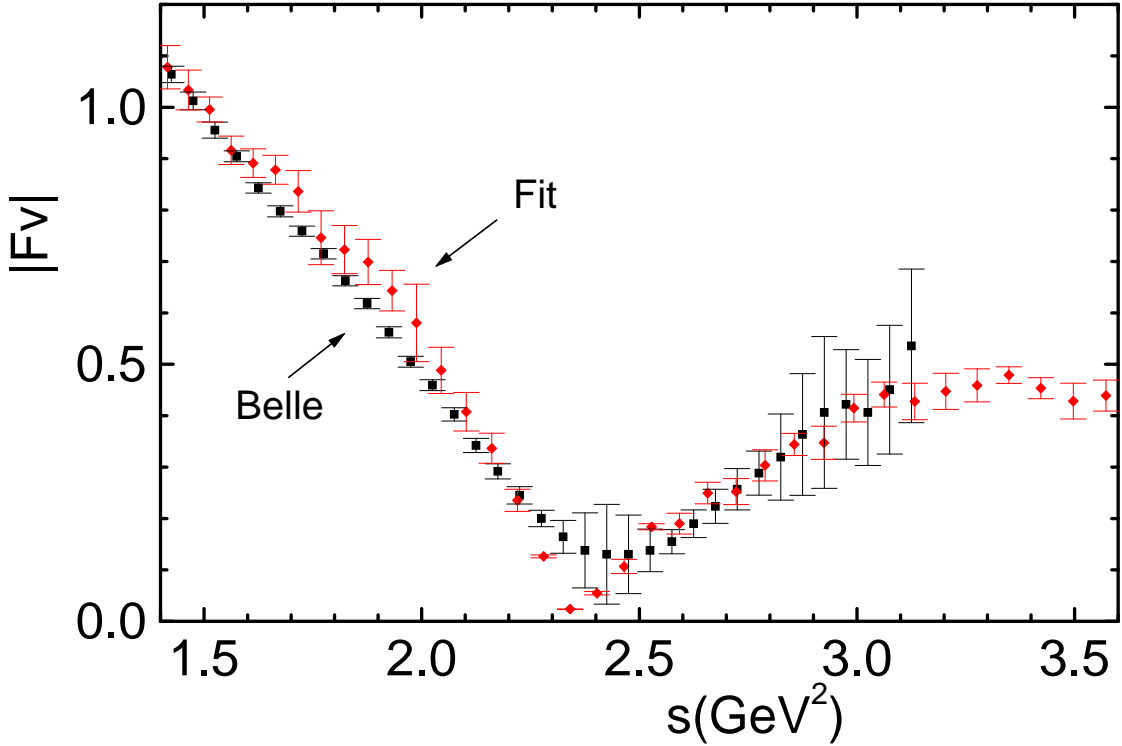}
\caption{Moduli of the Belle\cite{Belle} form factor (black)
and model results using CERN-Munich data \cite{Hyams} (red).}
\label{MOb}
\end{figure}
%

\section{conclusion}
\label{conc}

This work regards the structure of final state interactions in weak decays 
involving a two-pion channel.
With this purpose in mind, 
one studies the relationship between the weak 
two-pion vector form factor, measured in $\tau \rar \p \p \n$,
and free elastic $\p\p$ scattering.
On general topological grounds, contributions to $F_V^{\p\p}$
can be decomposed into two sets of Feynman 
diagrams, associated respectively with a primary weak kernel
and hadronic final state interactions(FSI).
One has chosen the latter to describe $T_{\p\p}^P$,  the observable $\p\p$ amplitude,
and this determines the weak kernel.
As the final pions can be produced either directly from $\cK_W$ or after rescattering, 
the decay amplitude is necessarily determined by the sum of these two 
kinds of contributions, as in Fig.\ref{FV}  and eq.(\ref{4}).
An important feature is that, in this sum, the relative weight of each term is unambiguously fixed by theory.
The scattering amplitude, in turn, can be written as the sum 
of a Dyson series, involving complex interaction kernels $\cK_{\p\p}^P$ and two-pion propagators 
$\Ob_{\p\p}^P$.
This allows both direct and FSI contributions to $F_V^{\p\p}$ to be incorporated   
into a compact dimensionless function $\T_{\p\p}^P$,
which is  determined directly from scattering data.
This means that all substructures
present in the $\p\p$ amplitude, such as resonances and inelasticities,
are automatically included into the calculation.  

The minimal prediction for $F_V^{\p\p}$, given by the product $F_\r\,\T_{\p\p}^P$, 
where $F_\r$ is the $\r$-meson-dominance  form factor,  
is largely dominant, as shown in Fig.\ref{FVThe}.
The more general expression $F_V^{\p\p} = \cK_W \, \T_{\p\p}^P $, 
eq.(\ref{5}), describes a conceptual separation  between weak kernel and 
final state interactions, which could easily be extended to heavy-meson decay processes.
Moreover, as both $F_V^{\p\p}$ and $\T_{\p\p}^P$ can be extracted from experiment, 
one can isolate empirical information about $\cK_W$,  
as indicated in Fig.\ref{KW}.
This has a practical implication because, if one chooses to fit $F_V^{\p\p}$ directly,
for consistency, one must  also fit the associated $\p\p$ scattering amplitude. 
The overlap between both observables is substantial, 
as can be inferred from Fig.\ref{FVThe}, 
and so is the amount of double fitting.
The decomposition proposed here suggests that the fitting process could be restricted  
just to $\cK_W$, which has a much simpler structure, owing to the absence of 
the two-pion cut.

Chiral perturbation theory is possibly the best theoretical
framework for treating the form factor  below 1 GeV.
As shown in Fig.\ref{MOa}, a $\T_{\p\p}^P$  based on the precise CGL chiral $\p\p$ amplitude\cite{CGL},
together with a single free parameter in $\cK_W$, fixed at $F_V G_V/F^2=1.20\,$,
gives rise to a very good prediction for $|F_V^{\p\p}|$ in this region.
Higher energies can also be encompassed reasonably well, as indicated by Fig.\ref{MO}, 
by means of simple model, discussed in Sect.\ref{weakmodel}.

In the study of hadronic heavy-meson decays, 
analyses of experimental data usually employ, simultaneously,
both non-resonant and resonant guess functions.
In terms of Feynman diagrams, the former amount to weak kenels,
whereas the latter include both the same weak kernels and 
meson-meson final state interactions.
However, in most works, non-resonant and resonant components
are parametrized independently,
whereas results presented here indicate that they must match precisely
and be written as {\em products}
of weak kernels by functions $\T\,$, suitably adapted to each
channel involved.
Theoretical tools of this kind could reduce the amount of fitting needed and prove to be more rewarding.
This possibility is presently being explored.

\section*{ACKNOWLEDGMENTS}
I thank Diogo Boito for a critical reading of an early 
version of this manuscript.

\appendix 
\section{two-meson propagator}
\label{propagator}

Results presented here are conventional and displayed for the sake of completeness. 
One deals with $P$ waves and the corresponding two-meson propagator is associated 
with the integral
\bea
&&
I_{aa}^{\m \n} 
= S\int \frac{d^4  \ell}{(2\p)^4}
\frac{\ell^\m  \ell^\n} {D_- \; D_+ },
\nn\\[2mm]
&& D_\pm = (\ell \sm q/2)^2 \pm  M_a^2 \;,
\label{A1}
\eea
where $S=1/2$ is the symmetry factor, 
and its Lorentz structure reads
\bea
&&
I_{aa}^{\m \n}  = g^{\m\n} \, A + q^\m q^\n \, B \;.
\label{A2}
\eea
Multiplying both eqs.(\ref{A1}) and (\ref{A2}), successively  by $2q_\m$ and by $g_{\m\n}$,   
using $2 q\cdot \ell = D_+-D_-$,  $\ell^2 = - \lp q^2/4 - M_a^2 \rp + (D_+ + D_-)/2$,
and equating results, one finds the conditions 
\bea
&& A + q^2 B = I_a/2 \;,
\nn\\[2mm]
&& 4 A + q^2 B = - (q^2/4 - M_a^2) \, I_{aa} + I_a \;,
\nn\\[2mm]
&& I_a = \int \frac{d^4  \ell}{(2\p)^4} \frac{1}{\ell^2 - M_a^2}\;,
\nn\\[2mm]
&&  I_{aa}
= S\int \frac{d^4  \ell}{(2\p)^4}
\frac{1} {D_- \; D_+ }\;,
\nn
\eea
which yield 
\bea
&&
I_{aa}^{\m \n}  = \lb g^{\m\n} - \frac{q^\m q^\n}{q^2} \rb 
\lb - \frac{1}{4} \lp q^2 - 4 M_a^2 \rp \;\frac{ I_{aa}}{3} + \frac{I_a}{6} \rb 
+ \frac{q^\m q^\n}{q^2} \, \frac{I_a}{2} \;.
\label{A3}
\eea
The integrals  in this result can be evaluated using dimensional techniques\cite{GL}
and read\cite{Sch}
\bea
&& I_a = - i\, \frac{M_a^2}{16 \p^2} \, \lb R + \ln \frac{M_a^2}{\m^2} \rb 
\nn\\[2mm]
&& I_{aa} = -\, \frac{i}{16 \p^2} \, \lb R + \ln \frac{M_a^2}{\m^2} + 1 \rb - i\, \Ob_{aa} \;,
\nn
\eea
where $R$ is a function of the number of dimensions $n$, which diverges in the limit $n\rar 4$\cite{Sch},
$\m$ is the renormalization scale, and $\Ob_{aa}$ a the regular function
which, for $q^2 \geq 4M_a^2$, has the form
\bea
&& \Ob_{aa}= -\frac{1}{32 \p^2} 
\lc 2 - \sqrt{\frac{q^2 \sm 4M_a^2}{q^2}} \; 
\ln\lb \frac{q^2- 2M_a^2 + \sqrt{q^2 (q^2\sm 4M_a^2)}}{2M_a^2}\rb
+ i\, \p\, \sqrt{\frac{q^2 \sm 4 M_a^2}{q^2}} \;\rc\;.
\label{A4}
\eea
In the renormalization process, the divergent factors $R$ are
replaced by undetermined constants.
However, there is no need to face this problem here, since one is concerned just
with on-shell contributions to the propagator, associated with 
its imaginary part.  
This corresponds to the approximation
\bea
&& I_{aa}^{\m\n} = \frac{i}{4}\, 
\lb g^{\m\n} - \frac{q^\m q^\n}{q^2} \rb\; \Ob_{aa}^P \;,
\nn\\[2mm]
&& \Ob_{aa}^P = -  \frac{i}{96 \pi} \, \frac{[q^2 \sm 4\,M_a^2]^{3/2}}{\sqrt{q^2}}  \;.
\label{A5}
\eea

\section{kernels and amplitudes}
\label{kernels}

The use of coupled channels in meson-meson scattering is traditional
and one just summarizes main results,
in order to set the notation, which is close to that
used in Ref.\cite{Hyams}. 
One employs three basic kernels, represented as follows:
$\cK_{11} \rar [\p\p \lrar \p\p]$,
$\cK_{12} \rar [\p\p \lrar KK]$ and
$\cK_{22} \rar [KK \lrar KK]$. 
They are defined as sets of proper Feynman diagrams,
in the sense that they cannot be split by cutting two 
on-shell meson lines only.
In the case of $P$-waves, they are proportional to $(t \sm u)$
and one writes $\cK_{ij}=(t \sm u)\,\cK_{ij}^P$.
The iteration of these kernels, using the two-meson propagators
given in App.\ref{propagator},
yields diagonal amplitudes, written as
\bea
T_{ii}^P= \frac{\cK_{ii}^P}{1 + \Ob_{ii}^P\,\cK_{ii}^P} \;,
\label{B1}
\eea
with $i=1,2$.
The full $\p\p$ elastic amplitude includes $\p\p-KK$ oscillations 
and has the typical coupled channel form, given by
\bea
&& T_{\p\p}^P = \frac{\cK_{\p\p}^P}{1 + \Ob_{\p\p}^P\,\cK_{\p\p}^P} \;,
\nn\\[2mm]
&& \cK_{\p\p}^P = \cK_{11}^P 
- \cK_{12}^P \lb 1 \sm \Ob_{KK}^P\, T_{22}^P \rb
\Ob_{KK}^P \, \cK_{12}^P \;.
\label{B2}
\eea
Using eq.(\ref{B1}), one also has
\bea
&& \cK_{\p\p}^P = \cK_{11}^P 
- \frac{\cK_{12}^P\, \Ob_{KK}^P \, \cK_{12}^P}
{1 + \Ob_{KK}^P\,\cK_{22}^P} \;,
\label{B3}
\eea
where $\cK_{\p\p}^P$ becomes complex above the $KK$ threshold.
This gives rise to the usual form
\bea
&&T_{\p\p}^P = 
\frac{\cK_{11}^P 
+ \Ob_{KK}^P ||\cK||}
{1 + \Ob_{\p\p}^P \, \cK_{11}^P + \Ob_{KK}^P \, \cK_{22}^P
+ \Ob_{\p\p}^P\,\Ob_{KK}^P \,||\cK||} \;,
\nn\\[2mm]
&& ||\cK|| = \lb \cK_{11}^P\,\cK_{22}^P \sm (\cK_{12}^P)^2\,\rb \;.
\label{B4}
\eea

In models based on chiral symmetry, the kernels $\cK_{ij}$
are given by a leading contact interaction\cite{GL},
supplemented by resonances\cite{EGPR}.
At low energies, below $\sqrt{s}\sim m_\r$, processes involving tree level
contact and resonance interactions provide good guidance for 
the magnitude of interactions.
In that region, kernels are real and,  
in the framework of $SU(3)$, one has the relationship
\bea
&& \cK_{11}^P = \rtw\;\cK_{12}^P = 2\, \cK_{22}^P \;.
\label{B5}
\eea

\section{function $\T_{\p\p}^P$}
\label{empf}

The evaluation of the form factor, eq.(\ref{5}),
requires the function 
$ \T_{\p\p}^P= 1/[1 + \Ob_{\p\p}^P\,\cK_{\p\p}^P]$.
The relevant kernel is obtained by inverting eq.(\ref{B2}),
and one has
\bea
&& \cK_{\p\p}^P = \frac{T_{\p\p}^P}{1 - \Ob_{\p\p}^P\,T_{\p\p}^P} \;.
\label{C1}
\eea
The amplitude $T_{\p\p}^P$ is related to its non-relativistic
counterpart $t_{\p\p}\,$, used in phase shift analyses, 
by\cite{GL}
\bea
T_{\p\p}^P &\!=\!& 
96\p \lb \frac{s}{\lp s \sm 4 M_\p^2\rp^3}\rb^{1/2} t_{\p\p} \;,
\label{C2}
\eea
with
\bea
t_{\p\p} &\!=\!& \frac{1}{2 i}\lb \eta\, e^{2i\d}-1 \rb 
= \frac{\tan \theta}{1- i\,\tan \theta} \;,
\nn\\[2mm]
\theta &\!=\!& \d -i \, \ln \sqrt{\eta} \;. 
\label{C3}
\eea
Using these results, together with eq.(\ref{A5}),  into eq.(\ref{C1}), one finds 
\bea
&& \Ob_{\p\p}^P\, \cK_{\p\p}^P 
= -  i\, \tan \theta \;,
\label{C4}
\eea
and hence,
\bea
&& \T_{\p\p}^P = \frac{1}{1 - i \, \tan\theta} \;,
\nn\\[2mm]
&& i\, \tan\theta = \frac{[\eta^2 -1] + i\,[2 \eta \sin 2\d ]}
{1+\eta^2 + 2 \eta \cos 2\d} \;.
\label{C5}
\eea
%



\begin{thebibliography}{99}

\bibitem{Cleo} S. Anderson et al. (CLEO Collaboration),
Phys. Rev. D ({\bf 61}), 112002 (2000).

\bibitem{Aleph} S. Schael et al. (ALEPH Collabotation),
Phys. Rep. {\bf 421}, 191 (2005).

\bibitem{Belle} M. Fujikawa et al. (BELLE collaboration),
Phys. Rev. D {\bf 78}, 072006 (2008).

\bibitem{th} For a comprehensive review with many references, see A. Pich, 
Progr. Part. Nucl. Phys {\bf 75}, 41 (2014).

\bibitem{IT} I. Bediaga, T. Frederico and O. Louren\c co,
Nucl. Part. Phys. Proc.{\bf 258-259}, 167 (2015);
J.H.A. Nogueira, I. Bediaga, A.B.R. Cavalcante,
T. Frederico and O. Louren\c co, e-Print: arXiv: 1506.08332 [hep-ph].

\bibitem{AI} A. C. dos Reis and I. Bediaga, private communication.

\bibitem{Hyams} B. Hyams et al (CERN-Munich Collaboration),
Nucl. Phys. B {\bf 64}, 134 (1973).


\bibitem{BR} P. C. Magalh\~{a}es, M. R. Robilotta,
K. S. F. F. Guimar\~{a}es, T. Frederico,W. S. de Paula,
I. Bediaga, A. C. dos Reis, 
and C.M. Maekawa,
Phys. Rev. D{\bf 84}, 094001 (2011). 


\bibitem{CGL} G. Colangelo, J. Gasser and H. Leutwyler,
Nucl. Phys. B {\bf 603}, 125 (2001).

\bibitem{GL} J. Gasser and H. Leutwyler,
Ann. Phys. {\bf 158}, 142 (1984);
Nucl. Phys. B{\bf 250}, 465 (1985).



\bibitem{EGPR} G. Ecker, J. Gasser, A. Pich and E. De Rafael,
Nucl. Phys. B {\bf 321}, 311 (1989).

\bibitem{SCP} J.J. Sanz-Cillero and A. Pich, Eur. Phys. J C {\bf 27}, 587 (2003).

\bibitem{Sch}  S. Scherer,  Adv. Nucl. Phys. {\bf 27}, 277  (2003).



\end{thebibliography}
\end{document}